\documentclass[conference]{IEEEtran}  
\IEEEoverridecommandlockouts
\usepackage{cite}
\usepackage[hyphens]{url} 
\usepackage{hyperref} 
\usepackage{amsmath,amssymb,amsfonts}
\usepackage{graphicx}
\usepackage{textcomp}
\usepackage{xcolor}
\usepackage{subcaption}
\usepackage{verbatim}
\usepackage{caption} 
\usepackage{float}
\usepackage{array}
\usepackage{algorithm}
\usepackage{algpseudocode}
 
\def\BibTeX{{\rm B\kern-.05em{\sc i\kern-.025em b}\kern-.08em
    T\kern-.1667em\lower.7ex\hbox{E}\kern-.125emX}}

\newcommand{\shb}[1]{\textcolor{purple}{\textsc{SB:} #1}}

\begin{document}

\title{Blockchain-based Trust Management in Security Credential Management System for Vehicular Network \\
}

\author{\IEEEauthorblockN{SangHyun Byun, Arijet Sarker\IEEEauthorrefmark{3}, Sang-Yoon Chang, Jugal Kalita}
\IEEEauthorblockA{
\textit{University of Colorado Colorado Springs, Florida Polytechnic University\IEEEauthorrefmark{3}}\\ 
\{sbyun,schang2,jkalita\}@uccs.edu, asarker@floridapoly.edu\IEEEauthorrefmark{3}}
}

\maketitle
\begingroup\renewcommand\thefootnote{\textsection}

\endgroup

\begin{abstract}
Cellular networking is advancing as a wireless technology to support diverse applications in vehicular communication, enabling vehicles to interact with various applications to enhance the driving experience, even when managed by different authorities. Security Credential Management System (SCMS) is the Public Key Infrastructure (PKI) for vehicular networking and the state-of-the-art distributed PKI to protect the privacy-preserving vehicular networking against an honest-but-curious authority using multiple authorities and to decentralize the trust management. We build a Blockchain-Based Trust Management (BBTM) to provide even greater decentralization and security. Specifically, BBTM uses the blockchain to 1) replace the existing Policy Generator (PG), 2) manage the policy of each authority in SCMS, 3) aggregate the Global Certificate Chain File (GCCF), and 4) provide greater accountability and transparency on the aforementioned functionalities. We implement BBTM on Hyperledger Fabric using a smart contract for experimentation and analyses. Our experiments show that BBTM is lightweight in processing, efficient management in the certificate chain and ledger size, supports a bandwidth of multiple transactions per second, and provides validated end-entities.

\end{abstract}

\vspace{0.1in}
\section{Introduction}
\label{intro}

The vehicular networking technology has the potential to significantly improve the connected vehicle with the cellular network that accesses the cloud services and shares the information. Typically, it is significant for Vehicle-to-Everything (V2X) and Vehicle-to-Vehicle (V2V) communication between vehicles, devices, and infrastructures to provide secure network and privacy because the continuous broadcast of Basic Safety Messages (BSMs) potentially reduce unimpaired vehicle accidents by 80\%~\cite{brecht2019security}. In this regard, the US Department of Transportation (USDOT) proposed a project called Security Credential Management System (SCMS) for V2X communication to support these security requirements of vehicular networking. Vehicles broadcast BSMs to support V2X safety applications such as autonomous vehicles or connected autonomous vehicle. The backbone of the autonomous vehicle system is the onboard intelligent processing capabilities using the real-time data collected through V2X communication and sensors such as cameras, RADAR, LIDAR, and mechanical control units. This data includes BSMs (sender’s time, position, and speed etc.), traffic flow, road, and network conditions, etc. The sending vehicle signs each BSM and the receiving vehicle verifies the signed BSM to prevent an attacker from injecting false messages.

In SCMS, a Public Key Infrastructure (PKI) is proposed to issue pseudonym certificates to vehicles and infrastructure devices for maintaining reliable communication between them by dividing the generation and provisioning process of those certificates among multiple organizations. However, a recent project has shown many limitations of SCMS such as short-lived pseudonym CA, non-guaranteed quality of service (QoS), short-lived infrastructure-to-vehicle (I2V), and unbounded channel access delay~\cite{furtado2018threat}~\cite{qayyum2020securing}. SCMS provides Trust Management (TM) via sophisticated processes, such as the Elector-based Root Management (EBRM) and pseudonym Certificate Authority (PCA). TM introduces multiple distributed sets of authorities to generate and govern the certificate chain and policy generation files (required to generate and revoke the certificates for other authorities). Our work focuses on the TM providing the Chain of Trust for the PKI and managing the policy for authorities. While TM is designed to be resilient against authority compromise and provides distributed management using a sophisticated structure~\cite{brecht2019security}, our work using blockchain provides better resiliency and avoids the single point of failure in SCMS.

\textbf{Contribution.}
Our work enhances the SCMS PKI based on distributed PKI management by introducing a blockchain for Blockchain-Based Trust Management (BBTM). BBTM replaces the Policy Generator with a single blockchain divided into functional smart contracts: 1) Global Certificate Chain File (GCCF) smart contract, which manages certificate addition and revocation, and securely shares GCCF among SCMS authorities; 2) Global Policy File (GPF) smart contract, which manages SCMS authorities' operations efficiently and securely. BBTM offers several security advantages, including improved decentralization in PKI, greater resilience against single points of failure, and enhanced transparency in certificate, policy, and CRL transactions. Additionally, BBTM simplifies the SCMS architecture by replacing the Policy Generator and reducing the SCMS manager's role in managing GCCF, policies, and CRLs through the blockchain protocol.

\textbf{Paper Organization.}
Section \ref{sec:background} describes the state of the art SCMS for vehicular networking PKI, blockchain-Based Root Management, and blockchain. Our work is based on the SCMS PKI system. Section \ref{ps} defines the contribution scope and threat model of our approach. Section \ref{sa} describes the design principle and system architecture of the proposed approach while Section \ref{sec:bbtmdesign} provides the actual design including transaction, functions of GCCF and GPF, and network setup in BBTM. Section~\ref{secur} and Section \ref{impl} presents the security analysis and the implementation details of the proposed design and the experimental result. Section \ref{rw} reviews the related work and Section \ref{conclusion} concludes the paper.

\section{Preliminaries} 
\label{sec:background}

We provide an overview of the SCMS in Section~\ref{sec:scms} and then focus on TM, management of authorities in the SCMS, since our research focuses on BBTM to advance the SCMS design.

\begin{table}[t]
\small
\centering
\begin{tabular} { |p{2cm} | p{5.5cm} | }
\hline 
\textbf{Acronym} & \textbf{Explanation}  \\\hline 
RCA & Root Certificate Authority \\\hline 
ICA & Intermediate Certificate Authority \\\hline
ECA & Enrollment Certificate Authority \\\hline
PG & Policy Generator \\\hline 
MA & Misbehaviour Authority \\\hline
RA & Registration Authority \\\hline
PCA & Pseudonym Certificate Authority \\\hline 
DCM & Device Configuration Manager \\\hline
OSP & Ordering Service Provider \\\hline 
GCCF & Global Certificate Chain File \\\hline
GPF & Global Policy File \\\hline
BBRM & Blockchain-Based Root Management \\\hline  
EBRM & Elector-Based Root Management \\\hline 
BBTM & Blockchain-Based Trust Management \\\hline 
SCMS & Security Credential Management System \\\hline 
\end{tabular}
\caption{Acronyms}
\label{table:acronyms}
\end{table} 

\subsection{Security Credential Management System} 
\label{sec:scms}
\begin{figure}[t]
\centering
\includegraphics[width=0.45\textwidth]{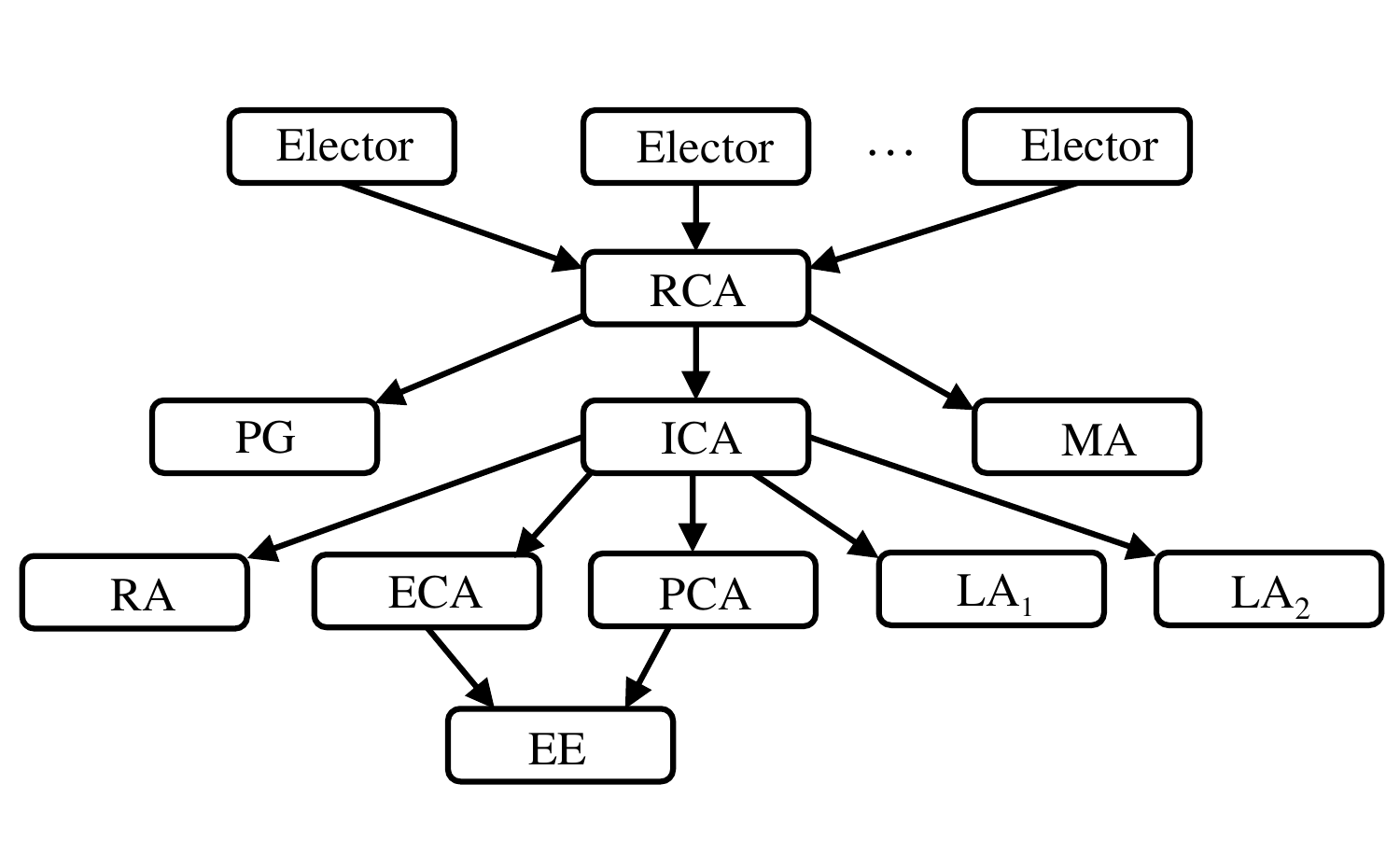}
\vspace{-2mm}
\caption{SCMS Authorities and Interactions} 
\label{fig:scmsarc}
\vspace{-5mm}
\end{figure}

The Security Credential Management System (SCMS)~\cite{brecht2018security}~\cite{whyte2013security} is a Public Key Infrastructure (PKI) that is used for issuing digital certificates to vehicles, infrastructure, and pedestrian/mobile devices involved in vehicle-to-everything (V2X) communications. The purpose of this PKI system is to enable secure and private communications between these different entities participating in the V2X network. The SCMS adopts a robust threat model to protect the privacy of vehicles. Beyond considering a standard adversary who may access the V2X communication channels when the digital certificates are being used, the SCMS also considers the possibility of an honest-but-curious adversary. This type of adversary would compromise the authority entities involved in the SCMS operations during the process of generating the digital certificates for vehicles and other entities. The SCMS is designed to safeguard against this stronger adversary model, not just the basic eavesdropping threat~\cite{chen2017protecting,brorsson2018guarding}. The strong threat model adopted by the SCMS motivates the use of a decentralized PKI (Public Key Infrastructure) architecture. This is to ensure a separation of the different SCMS authority entities so that a compromise of any single authority does not lead to a breach of vehicular privacy. Therefore, the SCMS provides a distributed PKI system involving multiple authorities, rather than a traditional centralized PKI which has a single Certificate Authority responsible for all certificate management tasks like signing, generation, and issuance. This decentralized, multi-authority approach is a key design choice of the SCMS to better protect the privacy of vehicles and other entities participating in the V2X communication network, even in the face of an honest-but-curious adversary compromising one of the authorities. Fig. ~\ref{fig:scmsarc} illustrates the different authority entities involved in the SCMS and the interactions between them. The lines connecting the entities represent the various interactions and operations that take place as part of the SCMS processes.

\subsection{Blockchain-based Root Management}
\label{ebrm}
In  SCMS, devices such as Registration Certificate Authority (RCA), Intermediate Certificate Authority (ICA), Pseudonym Certificate (PG), Management Authority (MA), Enrollment Certificate Authority (ECA), Root Authority (RA), Pseudonym Certificate Authority (PCA), Linkage Authority 1 (LA1), Linkage Authority 2 (LA2), and End Entity (EE) create a standard Public Key Infrastructure (PKI) hierarchy. This hierarchy maintains a chain of trust, which is an ordered list of certificates that allows the receiver to verify all certificates up to the top of the chain, among the authorities in SCMS and EE devices. Elector-based Root Management (EBRM)~\cite{sarker2021blockchain} and standard PKI hierarchy involve a group of electors who are responsible for overseeing and making decisions about the RCA. The electors use their self-signed certificates and follow a democratic voting process to add or remove RCA certificates or elector certificates. 
In BBRM, electors vote for adding or revoking the certificate of RCA or elector by signing an endorsement and then all endorsements are aggregated into the ballot ledger in the blockchain. A ballot ledger can be trusted by other authorities in the blockchain system where is a quorum of valid elector endorsements. This voting rule is defined in the GPF. There are four types of endorsement; adding RCA certificate, revoking RCA certificate, adding an elector certificate, and revoking an elector certificate. BBRM can track the history of the certificate because blockchain can record all of transactions of endorsement. In SCMS, SCMS manager conducts this voting process.

\section{Problem Statement}
\label{ps}

We describe the problem statement and the scope of the contribution in Section~\ref{subsec:problem_scope} and the threat model in Section~\ref{subsec:tm}. 

\subsection{Contribution Scope}
\label{subsec:problem_scope}

Although there are multiple authorities for SCMS, our focus is on TM consisting of whole authorities in SCMS. Securing the trust of the Chain is critical for PKI because it provides the trust relationship and security for SCMS, which in turn provides the certificates and the trust in keys to the vehicles for vehicular networking. Therefore, our work defends against the single authority in SCMS compromise by having a blockchain to replace the SCMS Manager divided into PG functions (aggregating the GCCF and managing the GPF) and securely to control the authorities through GPF. Our work also focuses on the blockchain to replace the SCMS Manager's role (whose primary role is to set and administer the SCMS policy) and reduce the authorities in SCMS since the smart contracts replace several functions in the SCMS that have previously been executed by the SCMS Manager.

\subsection{Threat Model}
\label{subsec:tm} 
In our threat model, we assume that adversaries cannot compromise the SCMS Manager, preventing them from controlling authority policies or manipulating the GCCF and GPF. We also assume that adversaries cannot break blockchain certificates using cryptographic methods like forging digital signatures or finding hash collisions. Within our BBTM Framework, we recognize that adversaries may monitor GCCF and GPF transactions as they are broadcast to SCMS authorities. We expect SCMS authorities to adhere to the GPF, which includes root management, addressing misbehavior in the MA, and managing the Certificate Revocation List (CRL), thus implementing accountability. Each authority has different access to smart contracts via the GPF. The distributed nature of the blockchain protects the integrity of stored data such as GCCF and GPF. We identify potential attacks, including blockchain tampering, information disclosure, privilege escalation, Denial of Service, Distributed Denial of Service, and Man-in-the-Middle attacks. Our approach effectively manages smart contract accessibility related to the addition or revocation of certificates and authority policy control.

\section{BBTM Design Principle and Architecture} 
\label{sa} 
In this section, we present the BBTM architecture and design principles by explaining the authorities and describing the advantages of applying blockchain to the SCMS. We also explain the rationale for using two blockchains to separate the distinct functionalities of the GCCF and the GPF, as opposed to using a single blockchain. The acronyms and their explanations used in our BBTM scheme are provided in Table \ref{table:acronyms}.


\subsection{Advantages of Using Blockchain for Trust Management}

BBTM applies blockchains for two functionalities: i) generating the GCCF from aggregating certificates to add or to revoke the certificate for each authority following the policy to replace the smart contract to make the SCMS simpler and rid the corresponding vulnerabilities, e.g., each authority being the single point of failure and ii) controlling the GPF to replace the smart contract to protect rule of authority and update new rule for authority.
Blockchain is appropriate for such purposes since the transactions within the ledger are immutable, transparent, automatic, and distributed. The immutability (after confirmation and the associated processing delay) enhances trust in transactions within the GCCF and GPF. This means it makes the transactions of generating GCCF through adding and revoking certificates more trustworthy, and it increases the reliability of transactions for generating GPF by adding or deleting authority policies. The transparency provides BBTM with better accountability for each smart contract issuing certificates from upper authorities such as Electors, RCA, ICA, etc and managing GPF by SCMS manager via PG. Blockchain protocol synchronizes the GCCF and GPF automatically and additionally. Finally, the blockchain concentrates on synchronizing data in the ledger through a distributed consensus protocol for GCCF and GPF. Note that SCMS is based on a distributed and decentralized network. BBTM replaces SCMS network divided into two functional authorities to decentralize the SCMS by avoiding the single authority  processing GCCF and GPF.


\subsection{Advantage of Using Two Blockchains Instead of One Blockchain}
BBTM provides two functionalities in SCMS (GCCF and GPF). We use two separate blockchains as opposed to one blockchain because different data formats are stored in each blockchain. Using two blockchains with two functionalities enables synchronization and causality control without extra mechanisms. For instance, If each authority who is in the process of each functionality sends a transaction in each separate blockchain, it can be more trustworthy than using one blockchain. In contrast, BBTM using two blockchains enables a clearer ordering service through the Ordering Service Provider in the permissioned blockchain, as it prevents forks from occurring.

\subsection{BBTM Structure and Entities}

\begin{figure}[t]
\centering
{\includegraphics[width=0.48\textwidth]{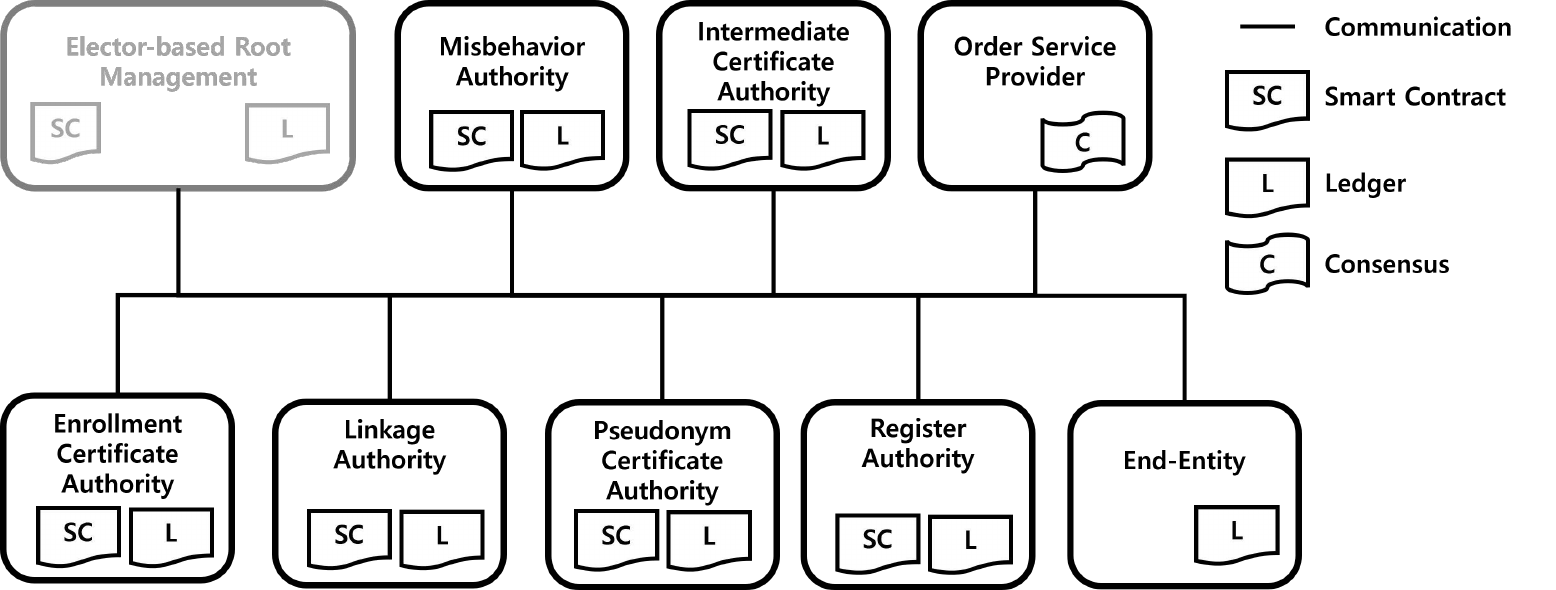}}
\caption{Blockchain-Based Trust Management (BBTM) 
}
\label{fig:BbTM}
\end{figure}

In this section, we describe the structure of BBTM and the authorities considered in BBTM for our proposed approach management. Our BBTM aims at two functions of SCMS manager to generate GCCF to aggregate all certificate processes in SCMS and to manage GPF. Fig.~\ref{fig:BbTM} shows an overview of the proposed BBTM architecture as well as the authorities considered in BBTM. The following authorities are part of the BBTM design: 

\begin{table}[t]
\small
\centering
\begin{tabular} { | m{1.2cm}| m{6cm} | }
\hline
\textbf{Notation} & \textbf{Description } \\\hline 
$S$ & Set of Entities in BBTM Blockchain \\\hline 
$PG$ & Policy Generator \\\hline 
$OSP$ & Ordering Service Provider \\\hline
$GCCF$ & Global Certificate Chain File \\\hline
$GPF$ & Global Policy File \\\hline
$CERT_{i}$ & Certificate of the Entity $i \in S$ \\\hline
$SIGN$ & Signed certificate algorithm \\\hline 
\end{tabular}
\caption{Notations and Variables}
\label{table:notations}
\vspace{-5mm}
\end{table}

\textbf{Elector:} The core element of BBTM, responsible for processing votes to add or revoke Root Certificates (RC) and Elector Certificates (EC) according to the GPF.

\textbf{Root Certificate Authority (RCA):} Another core component of BBTM, the RCA manages the addition and revocation of certificates for authorities such as ICA, MA, and PG, following the GPF.

\textbf{Policy Generator (PG):} Acts as the central manager within SCMS, controlling the GCCF and GPF, and aggregating certificates from all authorities.   

\textbf{Intermediate Certificate Authority (ICA):} A secondary Certificate Authority issued by the RCA, responsible for signing certificates for PCA, RA, ECA, and LA.  

\textbf{Pseudonym Certificate Authority (PCA):} Issues short-term pseudonym identification certificates for vehicles, with the ICA as its issuer.

\textbf{Register Authority (RA):} Validates and processes requests from vehicles through a certificate validation smart contract, issued by the ICA.

\textbf{Enrollment Certificate Authority (ECA):} Issues enrollment certificates for vehicles, containing real identification information such as owner, manufacturer, and color.   

\textbf{End-entity (EE):} Represents vehicles with read-only access to the blockchain, holding an Enrollment Certificate (EC) and a Pseudonym Certificate (PC). The addition and revocation of EE certificates are not covered in this paper, as they are not stored on the blockchain.

\textbf{Ordering Service Provider:} Manages transaction ordering for smart contracts related to GCCF and GPF, and broadcasts transactions to other BBTM authorities.  

Note that the GCCF structure in SCMS includes versioning, endorsements from EBRM, and certificates from entities such as Elector, RCA, ICA, LA, and PCA~\cite{GCCFBenedikt}, while the GPF structure consists of entities, rules, and statuses (alive or dead).  

All components in the Blockchain-Based Trust Management (BBTM) network communicate through a secure channel, as shown by the solid line in Fig.~\ref{fig:BbTM}. In our design, all SCMS components utilize a distributed ledger to represent both the GCCF and GPF, containing historical transactions related to certificate management and policy actions. Blocks aggregate transactions before being added to the blockchains, enabling tracking of certificate generation and revocation.

The GCCF blockchain supports two roles, allowing authorized SCMS members to either read or write via smart contracts. Authorized members receive new transaction blocks from the Ordering Service Provider, while smart contracts manage functions like adding, revoking, and validating certificates. The GPF blockchain restricts access to the PG for adding or deleting policies, with other authorities having read-only access. Consensus algorithms facilitate agreement among authorities on transaction content and order within the BBTM blockchain. We use a distributed consensus protocol designed for consortium blockchains, avoiding less efficient protocols intended for permissionless blockchains.

\captionsetup[table]{skip=12pt}
\begin{table*}[hbt!]
\small
\centering 
{\begin{tabular}{ |l|m{14.5cm}| } 
\hline
\textbf{Name} & \textbf{Description} \\\hline
Version & This attribute contains the version number of the certificate and helps to keep track of the change in the certificate format.  \\\hline
Serial Number & This attribute serves as a unique identifier to distinguish among certificates. \\\hline 
Subject Name* & This field provides information about the entities for which certificate addition or revocation is required, e.g., elector, ICA, MA, and PG, etc. \\\hline 
Issuer Name & This attribute contains the information of the elector who creates and signs the certificate. \\\hline 
Subject Public Key* & This field provides information about the public key of the entities for which certificate addition or revocation is required, e.g., elector, ICA, MA, and PG, etc. \\\hline 
Subject Unique ID* & This attribute contains the unique identifier value of the entities for which certificate addition or revocation is required, e.g., elector, ICA, MA, and PG, etc. \\\hline
Issuer Unique ID & This field contains the unique identifier value of the elector, who creates and signs the certificate.  \\\hline
Validity Period & This attribute holds two values: 1) valid from (determines the date when the certificate becomes valid) and 2) valid to (determines the date after which the certificate is expired). \\\hline
Digital Signature & This attribute contains the digital signature of the elector who creates and signs the certificate. \\\hline 
Algorithm & This field determines the name of the hashing algorithm and the signing algorithm used in the certificate.\\\hline
Function Type & We have two \textit{certificate} functions: i) \textit{AddCert}, and ii) \textit{RevokeCert}, This field contains the name of the specific function. \\\hline
 
\end{tabular}} 

\caption{Certificate Format using blockchain. The Asterisks (*) are the fields that are dependent on the function type and, more specifically, on whether it is for SCMS-authority certificate generation/revocation}

\label{fig:certificateformat}
\vspace{-6mm}
\end{table*}

\section{BBTM Design Principle }
\label{sec:bbtmdesign}
In this section, we describe the design of the network setup, functionalities, and transactions of GCCFL and GPF used in BBTM. We also describe the decentralized trust management for SCMS in BBTM.  

\subsection{Notations and Variables}
Notations and variables are explicitly defined in Table \ref{table:notations}. $S$ is the set of entities used in BBTM. The set of the entities involved in BBTM is $S$ =\{all authorities in SCMS\}. For any entity is $i$, where $i$ $\in$ $S$, $CERT_{i}$ defines the certificate of the entity $i$. We define Policy Generator, Order Service Provider, Global Chain Certificate File, and Global Policy File as $PG$, $OSP$, $GCCF$, and $GPF$ respectively. $SIGN$ defines the certificate sign algorithm.

\subsection{Network Setup} 
In this subsection, we outline the design of the BBTM network setup, establishing two channels: 1) the System Channel dedicated to storing consortium configuration, and 2) the Application Channel designed for sharing ledger and chaincode among channel members. At this stage, the initial set of all authorities is created, and credentials for them are generated on their respective machines.  

\begin{algorithm}[t]
\caption{Adding and Revoke Certificate}\label{al:addcert}
\begin{flushleft}
\hspace*{\algorithmicindent} \textbf{Input:} $CERT_i$, $CERT_i[Arguments]$, $CERT_{PG}$ \\
\hspace*{\algorithmicindent} \textbf{Output:} $GCCF$  
\end{flushleft}
\begin{algorithmic}[1]
\If {$Adding \& Verifying$ $CERT_i$ and $CERT_i[Arguments]$}
\State $CERT_i$ sign $CERT_i[Arguments]$
\State $GCCF$ $\gets$ signed $CERT_i[Arguments]$
\State $\textbf{return}$  $Updated GCCF.Function type$
\Else 
\State $\textbf{return}$ $Not Adding Verify$
\EndIf
\If {$Revoke \& Verifying$ $CERT_{PG}$}
\State $GCCF$ $\gets$ signed $CERT_i[Arguments]$
\State $\textbf{return}$ $Updated GCCF.Function type$
\Else 
\State $\textbf{return}$ $Not Revoking Verify$
\EndIf
\end{algorithmic}
\end{algorithm} 

\begin{algorithm}[t]
\caption{Validate Certificate}\label{al:valcert}
\begin{flushleft}
\hspace*{\algorithmicindent} \textbf{Input:} $CERT_{S}$,\\
\hspace*{\algorithmicindent} \textbf{Output:} $BOOL$  
\end{flushleft}
\begin{algorithmic}[1]
\If {$Verifying$ $CERT_S$ through $GCCF$}
\State $\textbf{return}$ $Success$
\Else 
\State $\textbf{return}$ $Not Verify$ 
\EndIf
\end{algorithmic}
\end{algorithm} 
\setlength{\textfloatsep}{1mm}
\setlength{\floatsep}{1mm}

\textbf{1) System Channel:} In this step, the genesis block of the system channel is generated through the Order Service Provider (OSP) registration and consortium Member Registration procedures. In the OSP Registration procedure, the configuration of the ordering service is established by incorporating all authorities and consensus algorithms, where is used by the permissioned consensus algorithm. The Consortium Member Registration procedure defines a consortium member by incorporating the certificate of the authorities. 

\textbf{2) Application Channel:} The Application channel is established through both Consortium Member Policy and Consortium Member Setting. In Consortium Member Policy, the guidelines for consortium members, such as the privileges for reading and writing in the consortium blockchain, are outlined. In Consortium Member Setting, information about consortium members, specifically in the context of BBTM, is included all authorities. The genesis block of application channel (GCCF and GPF) is generated using them and then it is broadcasted to all authorities from OSP. 

\begin{algorithm}[t]
\caption{Adding and Revoke Policy}\label{al:addpolicy}
\begin{flushleft}
\hspace*{\algorithmicindent} \textbf{Input:} $CERT_{PG}$, $Policy{live}$ \\
\hspace*{\algorithmicindent} \textbf{Output:} $GPF$  
\end{flushleft}
\begin{algorithmic}[1]
\If {$Adding \& Verifying$ $CERT_{PG}$}
\State $Policy.status = alive$
\State $\textbf{return}$ $Updated GPF$
\EndIf
\If {$Revoke \& Verifying$ $CERT_{PG}$}
\State $Policy.status=death$
\State $\textbf{return}$ $Updated GPF$
\EndIf
\end{algorithmic}
\end{algorithm} 

\subsection{Functionalities of GCCF and GPF} 
There are total five functions; 1) AddCert, 2) RevokeCert, 3) ValidateCert, 4) AddPolicy, and 5) RevokePolicy. The functions (AddCert, RevokeCert, and ValidateCert) within GCCF can collectively be referred to as PKI-signed, revoked, and validated functions. The AddCert belongs to the signed certificate process from a high-level authority in Algorithm~. For instance, RCA signs ICA, MA, and PG. The RevokeCert belongs to the revocation certificate process that the SCMS manager updates the message of a revoked certificate in Algorithm~\ref{al:addcert}. Transactions are submitted through two functions (AddCert and RevokeCert) utilizing the same payload format, indicating the same certificate format. However, the classification is divided into two types to differentiate between the distinct objectives of the function, specifying whether it is intended for adding or revoking a certificate. The certificate format adheres to the X.509 standard, incorporating an extended field labeled \emph{Function Type}. The description of the certificate field is defined in Table ~\ref{fig:certificateformat}. The ValidateCert function is used to validate certificates through reading transactions in blockchain. The pseudocode of ValidateCert function is shown in Algorithm~\ref{al:valcert}. Each transaction is associated with a key value, which serves as the index entry for that transaction. To retrieve the certificate information of a specific transaction, a search is conducted in the ledger using the transaction's index entry. If it cannot verify certificate, the error message is displayed; otherwise, a verified success message is displayed.    

The functions (AddPolicy and RevokePolicy) within GPF can collectively be referred to as managing policy file of all authorities in SCMS. The AddPolicy function is governed by the PG under the SCMS manager, while RevokePolicy also corresponds to the PG rules set by the SCMS manager. Transactions are submitted using two functions that contain the same payload format meaning the same policy file format including entities, rules, and statutes (alive and death). The pseudocode of AddPolicy and RevokePolicy are shown in Algorithm~\ref{al:addpolicy}. Note that rules in GPF are global configuration information~\cite{brecht2018security}.

\section{Security Analysis}
\label{secur}
In this section, we analyze the feasibility and viability of the threats and attacks outlined in our threat model.

\begin{figure}[t]
\centering
{\includegraphics [width=0.4\textwidth]{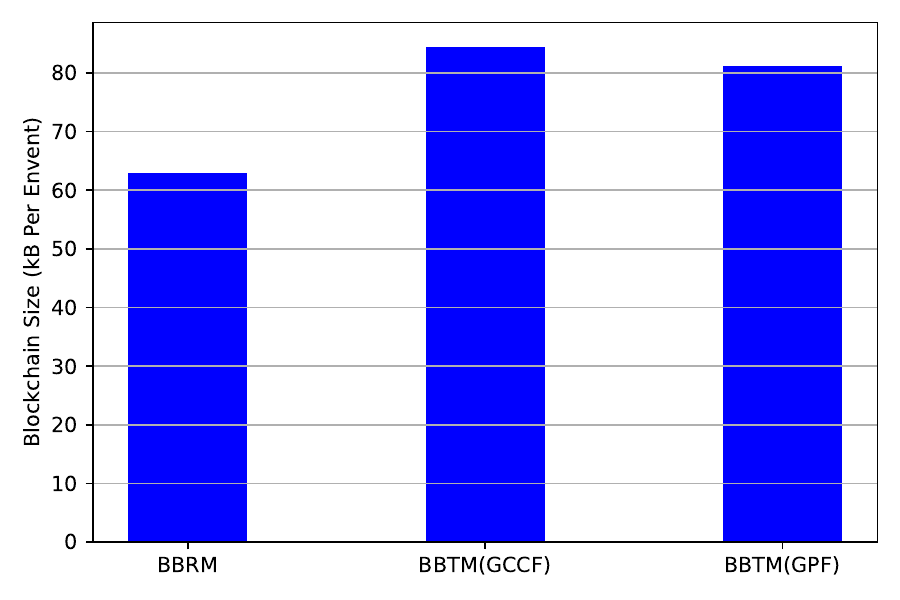}}
\caption{Blockchain Size 
}
\label{fig:bsize}
\end{figure}

\textbf{Privacy End-entity Preservation:}
In our proposed approach, the objective of encrypting the response message with the public key of a pseudonym certificate is to ensure that no unauthorized entities can access sensitive information. This encryption process enhances data confidentiality and protects the privacy of the communication, safeguarding it from potential threats and ensuring that only intended recipients can decrypt and access the information.

\textbf{Malicious or Compromised End-entity or authority:} The proposed architecture is designed to withstand the following attacks mentioned in our threat model:

Blockchain Attacks: Blockchain networks are generally regarded as highly secure and resistant to unauthorized manipulation because no single entity has control over all the authorization processes. To successfully launch an attack, an attacker would need to control a significant portion of the blockchain network, which is typically very difficult to achieve.

Man-in-the-Middle Attacks: The Trust Management platform, combined with a consortium blockchain, can securely store the trust certificate chain. This makes it extremely difficult for attackers to steal or alter these certificates, thereby protecting the integrity of the communication.

Denial-of-Service (DoS) and Distributed Denial-of-Service (DDoS): While PKI and TLS do not inherently prevent DoS or DDoS attacks, the shared blockchain ledger enables other nodes to maintain the BBTM system even in the event of an attack. If any nodes are compromised, they can recover data from their peers upon restoration, ensuring both data integrity and availability.



\section{Evaluation}
\label{impl}

\captionsetup[table]{skip=12pt}
\begin{table*}[hbt!]
\small
\centering 
{\begin{tabular}{ |p{2cm}|p{1.6cm}|p{1.6cm}|p{2cm}|p{2.5cm}|p{2.5cm}|p{2cm}| } 
\hline
 & \textbf{RAM (\%)} & \textbf{CPU (\%)}& \textbf{BLockchain size (kB)}& \textbf{Thoughput(Tx/s)}& \textbf{Thoughput(kB/s)}& \textbf{Transaction latency (s)}\\\hline
BBRM~\cite{sarker2021blockchain} & 31.80 & 4.2 & 62.8772 & 2.7926421 & 15.140614757501 & 358.658 \\\hline
BBTM(GCCF) & 18.80 & 5.8 & 84.3912 & 2.142140 & 13.58842377 & 624.714 \\\hline
BBTM(GPF) & 18.80 & 5.8 & 81.131 & 2.55033 & 14.2740025815 & 621.341 \\\hline
\end{tabular}} 
\caption{Blockchain Performance}
\label{tlb:Block}
\end{table*}

\subsection{Experimental Setup}
In this section, we outline the implementation details for BBTM. We developed a BBTM prototype using Hyperledger Fabric (version 2.0) to perform transactions on Google Cloud servers. This Proof-of-Concept (PoC) is essential for enabling vehicle communication with RA via cellular networks while driving. The experiment is conducted on a Google Cloud servers running Ubuntu, with a laptop positioned in a vehicle. We utilize LTE and 5G networks from a major mobile operator in the USA to facilitate communication between the vehicle and RA.  
\subsection{Experimental Results}
In this section, we show the analyses of BBTM and our system to show that it is appropriate for the SCMS. We analyze the transaction scalability in relation to the SCMS, blockchain size, resource consumption, throughput, and transaction time, as well as communication latency between vehicles and RA on 5G and LTE networks.

\subsubsection{Transaction Scalability and Blockchain size}
BBTM is for GCCF and GPF management within SCMS, meaning that vehicle certificate is out of our scope. Therefore, the application does not require high transaction scalability. SCMS is designed to manage the lifecycle of authorities' certificates, which can range from 1 to 12 years~\cite{TimelineBenedikt}.

In this section, we examine transaction scalability and the blockchain size growth rate in kB per event. Each event corresponds to a function of the Global Certificate Chain File (GCCF) and Global Policy File (GPF). BBTM emphasizes trust management in SCMS and demonstrates transaction scalability that is comparable to or lower than most other blockchain implementations. Its scalability aligns with the default capabilities of Hyperledger Fabric, which supports a ledger growth rate of 20,000 transactions per second at 2.9 kB per transaction, equating to 58,000 kB/s or approximately 1.829××10121012 B/year. As shown in Fig.~\ref{fig:bsize}, the blockchain size growth rates for BBTM(GCCF) and BBTM(GPF) are 1.3 $\times$ $10^{-7}$ and 14 $\times$ $10^{-7}$ kB/year respectively. Notably, the per-event scalability of BBTM is about ten orders of magnitude smaller than the capacity supported by Hyperledger Fabric.

\subsubsection{Transaction Throughput}

Transaction throughput is a key metric for evaluating the speed of a blockchain system, representing the rate at which valid transactions are committed by the network over a specific time period. In our study, we assess transaction throughput for five functions in BBTM, as these transactions can alter the blockchain state. Throughput is measured in transactions per second (Tx/s) and kilobytes per second (kB/s). We calculate transaction latency by averaging results from five iterations, each generating 1,000 transactions. As shown in Table~\ref{tlb:Block}, the transaction throughput for BBTM (GCCF) and BBTM (GPF) is approximately 2.1 and 2.5 Tx/s, and 13 and 14 kB/s, respectively. The differences in transaction throughput between BBTM (GCCF) and BBTM (GPF) are due to variations in blockchain size, with BBTM having a smaller size than GCCF and GPF, allowing events to complete in just a few seconds.

\begin{figure}[t]
\centering
{\includegraphics [width=0.41\textwidth]{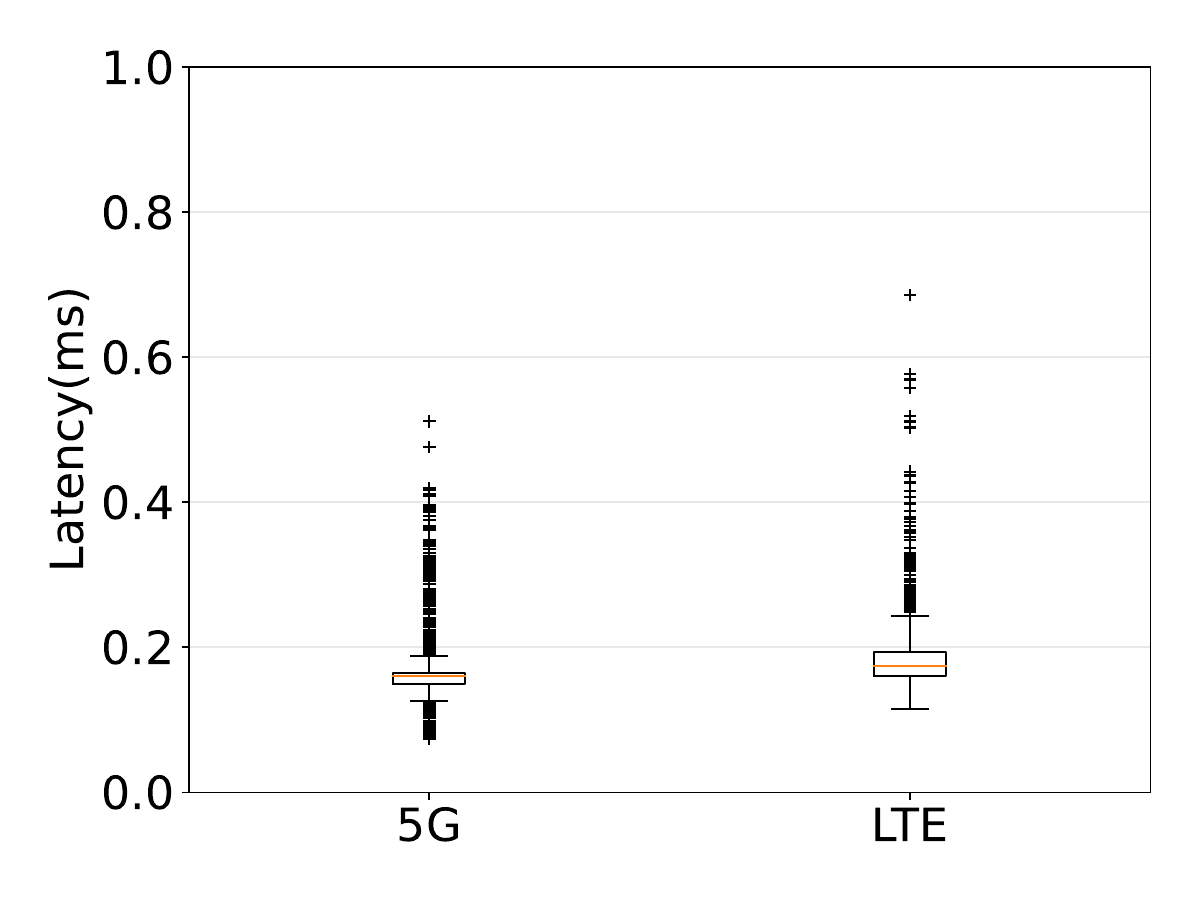}}
\caption{Communication latency between vehicle and RA, executed ever cellular networks (5G and LTE) during driving 
}
\label{fig:lat}
\end{figure}

\subsubsection{Resource Consumption}
\begin{figure}[t]
\centering
{\includegraphics [width=0.4\textwidth]{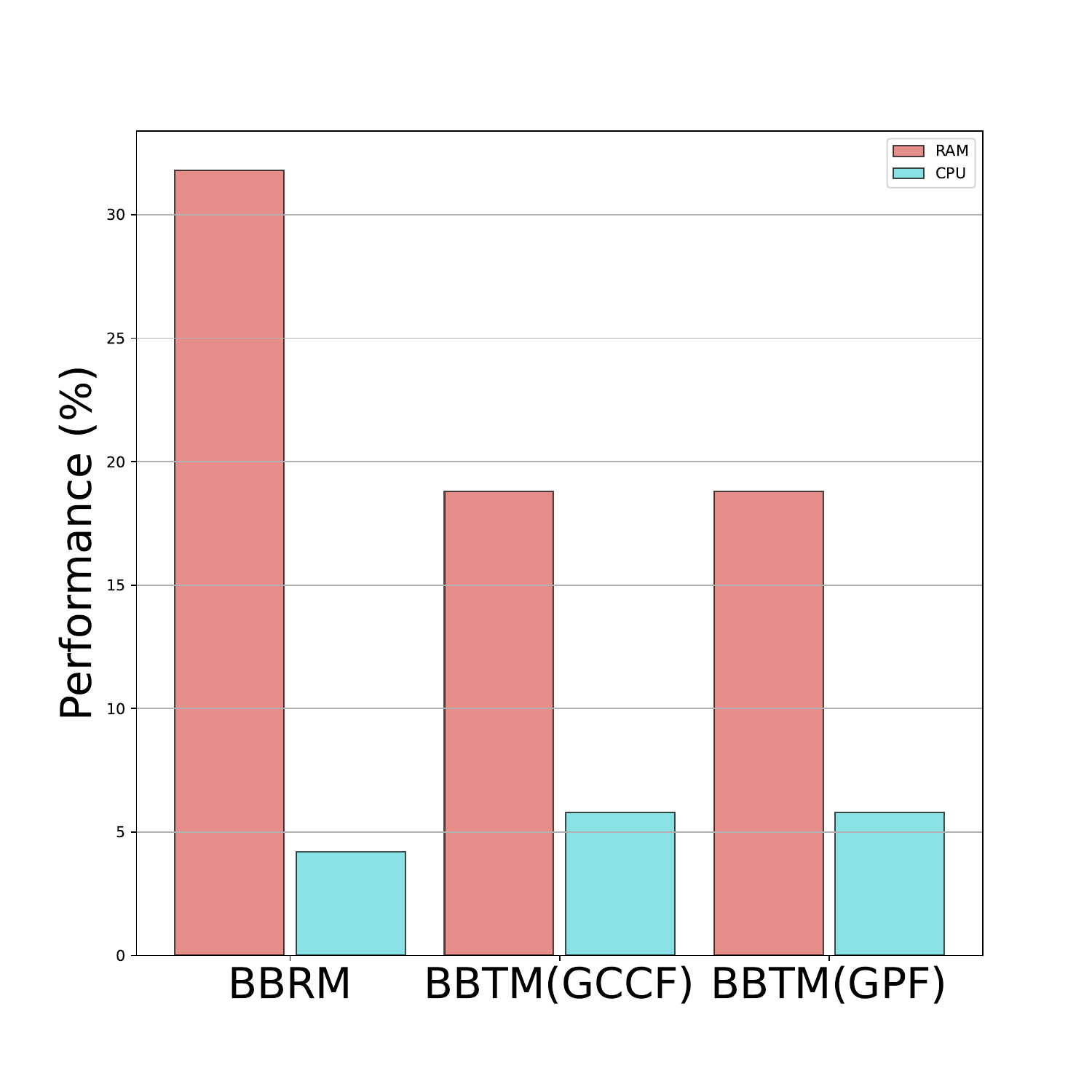}}
\vspace{-6mm}
\caption{Blockchain Performance 
}
\label{fig:per}
\end{figure}

We have also calculated the resource consumption (CPU usage, RAM) for running the blockchain for two different scenarios. All authorities can host various applications, each responsible for executing all the functions associated with a BBTM (GCCF) and BBTM (GPF), thereby consuming the required resources.  This blockchain application will indeed consume resources, and as a result, the measurement can provide insights into the configuration requirements for running all authorities. We have 4 GB memory for each of Google Cloud servers. We have analyzed the percentage of RAM utilized in each machine for two distinct scenarios, such as BBTM (GCCF) and BBTM (GPF). It is noticeable that the percentage of RAM usage has also decreased by more than 2x for BBRM, as it constitutes 31.8\% of the total 2048 GB RAM, while BBTM utilizes 18.8\% of the 4056 GB RAM. However, the percentage of CPU did not change in all scenarios. The results are shown in Fig.~\ref{fig:per}.

\subsubsection{Transaction and Communication Latency}
"In blockchain systems, transaction latency refers to the time required for a transaction to be accepted across the network. This includes the period from when a client initiates a transaction to its widespread availability. Specifically, transaction latency encompasses the time taken for a client to submit a transaction to peers, the execution of that transaction, the return of responses to the client, and the subsequent dissemination of the transaction and responses to the Ordering Service Provider (OSP), which then delivers them to all peers. We calculate transaction latency by averaging results from five iterations, each involving the generation of 1,000 transactions. As shown in Table~\ref{tlb:Block}, the transaction latencies for BBTM (GCCF) and BBTM (GPF) are 624.714s and 621.341s, respectively. Our analysis indicates that cloud-based services contribute to increased transaction latency compared to BBRM.

In our study, communication latency is defined as the total time required to verify the certificate from the EE to the RA. We evaluate communication latency using several statistical measures: the median represents the average latency; the Interquartile Range (IQR) indicates connectivity stability; whiskers show less stable connectivity; and crossed-outlier points highlight instability. The combination of the median and IQR also illustrates latency skewness. As shown in Fig.~\ref{fig:lat}, the medians for communication latency are 0.159 ms for 5G and 0.173 ms for LTE. The IQR for 5G ranges from 0.149 to 0.164 ms, while for LTE, it ranges from 0.160 to 0.193 ms. Out-of-range latencies are 0.125 to 0.188 ms for 5G and 0.114 to 0.243 ms for LTE.
\section{Related Work}
\label{rw}
In this section, we describe the related work of blockchain for decentralized PKI and the application of certificate management for vehicular networks.

\textbf{Blockchain for Decentralized PKI:} Blockchain is used to create a decentralized Public Key Infrastructure (PKI), improving resilience against central authority failures. In papers ~\cite{plessing2020revisiting, axon2016pb}, a blockchain-based PKI (BB-PKI) implemented via a CertCoin fork enhances anonymity by separating user identity from certificates. The paper~\cite{sermpinis2021detract} introduces DeTract, which enables web servers to independently generate X.509 certificates using uPort, an Ethereum-based identity platform for managing certificate transactions. Additionally, this paper~\cite{toorani2021decentralized} presents a blockchain-based Web of Trust (WoT) system that eliminates the need for a centralized Certificate Revocation List (CRL), while this paper~\cite{adja2021blockchain} improves X.509 validity mechanisms based on distribution points to reduce time delays. Finally, this paper~\cite{turan2024semi} presents SemiDec-PKI, a blockchain-based infrastructure that enhances fault tolerance and security for various certificate types by integrating a Web of Trust with a centralized model.

\textbf{Application of certificate management for Vehicular Networking:} Application of certificate management is applicable for blockchain to enhance security aspect. In this paper\cite{khan2020accountable}, the Accountable Credential Management System (ACMS) enhances vehicular communication security by transparently managing certificates and ensuring trustworthiness without relying on traditional certificate revocation lists. This paper~\cite{byunx2023privacy} presents the challenge of establishing trust between these two separate PKIs for vehicular networks. In this paper~\cite{byun2024secure}, there is multi-authority management between the federated learning server and SCMS for the vehicular network. In \cite{lei2020blockchain}, the author describes the growing importance of security and privacy in Vehicular Communication Systems (VCS) within the Intelligent Transportation Systems (ITS).

\section{conclusion and future work}
\label{conclusion}
The Security Credential Management System (SCMS) is an advanced Public Key Infrastructure (PKI) for vehicular networking, involving multiple authorities and distributed operations. It utilizes a distributed PKI system to ensure privacy-preserving management. Acknowledging its critical role in vehicular communications, we introduce Blockchain-Based Trust Management (BBTM) as an enhanced version of SCMS PKI management. BBTM builds on the existing functionalities of the Global Certificate Chain File (GCCF) and Global Policy File (GPF) within the Policy Generator (PG) of SCMS, enhancing security and integrity through blockchain technology, which offers high transaction integrity and resilience against compromises. Our experiments with Hyperledger Fabric demonstrate that BBTM achieves suitable transaction performance and execution overheads for GCCF and GPF. This work extends the current SCMS design for vehicular networks, with future directions focusing on developing a framework for various advanced PKI designs that rely on multiple authorities for distributed operations, particularly those emphasizing privacy and compromise resilience.


\bibliographystyle{IEEEtranS}
\bibliography{bibliography}

\end{document}